\documentclass[manuscript]{aastex}

\shorttitle{Photometric Study of MBM 12}
\shortauthors{Cohen, Herbst \& Williams}

\begin{document}

\title{A Photometric Study of Stars in the MBM 12 Association}

\author{William Herbst \& Eric C. Williams}
\affil{Astronomy Department, Wesleyan University, Middletown, CT 06459}
\and
\author{Wendy P. Hawley}
\affil{Astronomy Department, Wellesley College, Wellesley, MA 02481}

\begin{abstract}

We have monitored four fields containing nine previously identified members of the MBM 12 association to search for photometric variability and periodicity in these pre-main sequence stars. Seven of the nine are found to be variable and definite periodicity (of 1.2, 2.6 and 6.2 days) is found for three of them, including the classical T Tauri star LkH$\alpha$ 264. Two other members are possibly periodic but each requires confirmation. In addition, a ``field" star that is associated with the X-ray source RX J0255.9+2005 was discovered to be a variable with a period of 4.2 days. Our results indicate that the photometric variability characteristics of the known MBM 12 association members are typical of  what is found in $\sim$few My old stellar groups such as IC 348, supporting arguments for a similar age. In particular, there is a mix of periodic and non-periodic variables with typical amplitudes (in Cousins I) of  0.1-0.5 mag, in addition to a small number of larger amplitude variables. The periods, as a group, are somewhat shorter than in IC 348 but when allowance is made for the known dependence of period on mass in pre-main sequence stars the difference may not be significant. Our data confirm and illustrate the value of photometric monitoring as a tool for identifying likely association members and for studying rotation in  extremely young stellar groups. 

\end{abstract}


\section{Introduction}

Photometric variability is a well known characteristic of young, low mass stars. Irregular variations of up to several magnitudes are seen as a result of accretion and, perhaps, occultation events within the dusty and gaseous disks surrounding classical T Tauri stars (CTTS). This behavior can persist for as long as $\sim$10 My as the example of TW Hya \citep{ab} illustrates. One also sees variations of typically a few tenths of a magnitude or less in weak-line T Tauri stars (WTTS) that may be largely or entirely attributed to cool (magnetic) spots on the stellar surfaces \citep{h94}. If the spot patterns persist for many rotations it is often possible to derive rotation periods for the stars by monitoring their photometric variations. Recent accounts of this type of work and its value have been given for the Orion Nebula cluster by \citet{hbj}, for the larger Orion association by \citet{s99}, \citet{chs} and \citet{r01}, for the young cluster IC 348 by \citet{chw} and for NGC 2264 by \citet{lm03}.

MBM 12 \citep{m85} is a nearby complex of dark clouds which has given birth to a small number of relatively low mass stars \citep{l01,o03} known as the MBM 12 association, or MBM 12A. The distance was originally thought to be only 65 pc \citep{h00a,h00b}, which would make MBM 12 the closest known star forming cloud. Recent studies, however, indicate much larger values near 300 pc \citep{l01, sckl, a02}. At the larger distance, the stars must be well above the main sequence and, therefore, extremely young ($\sim$1-5 My). The recent discovery of an edge-on disk around an association member \citep{j02,c02} supports this view, as does the detection of millimeter wavelength radiation from some members \citep{i03}. Our interest in the association came from its inclusion on a list of possible targets for a Space Interferometry Mission (SIM) satellite study to detect extra-solar planets by astrometric means \citep{b03}. While it may be too distant for that purpose, we have carried out a photometric study of nine of the twelve known members in four fields and report the results here. They provide some new insight into the nature of MBM 12A as well as some information on its members, including several rotation periods.

\section{Observations and Data Reduction}

The observations were obtained on $\sim$30 nights between 8 Nov 2001 and 8 March 2002 with a 1024$\times$1024 Photometrics CCD attached to the 0.6m telescope at Van Vleck Observatory, located on the campus of Wesleyan University.  The field of view is 10.2$\arcmin$ on a side and the size of a pixel is 0.6$\arcsec$.  Four fields (designated A, B, C and D) were chosen to include the largest number of MBM 12A members possible from the list of twelve given by \citet{l01}. On each clear night, a sequence of 5 one-minute exposures was taken through the Cousins {\em I} filter for each field, as well as twilight flats, bias frames, and dark frames.  These were combined into an image of effectively 5 minutes duration but with a greater dynamic range than would have been possible with a single exposure. Also, any effects of non-uniform tracking by the telescope are removed by this procedure. Preliminary reductions were done with standard IRAF tasks. Aperture photometry was performed on all uncrowded stars clearly visible on the images using the APPHOT package in IRAF.  An aperture radius of 7.5 pixels was adopted and an annulus with inner and outer radii of 10 and 15 pixels respectively was used to determine the sky level.  

\begin{figure}
\plotone{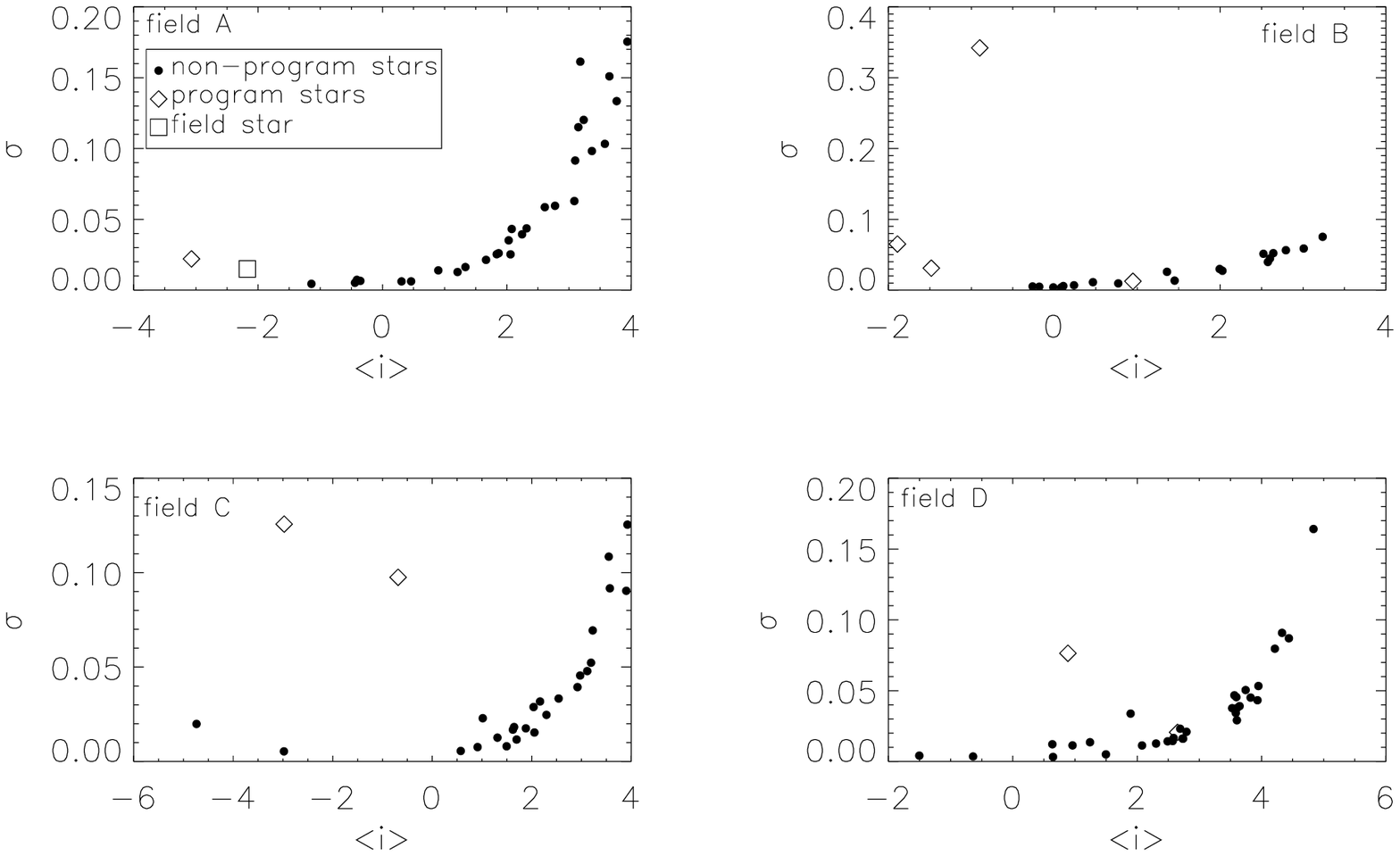}
\caption{Standard deviations ($\sigma$) for each star plotted as a function of average differential instrumental magnitude, <i>, based on the scatter in the photometry over the observing season. The four fields are plotted separately since different comparison stars are used. Known members of MBM 12A are indicated by the diamonds and the one variable field star detected is plotted as a square. The extremely bright star in field C is close to the saturation limit on many images and we believe that it is probably not actually variable in spite of the large $\sigma$.}
\label{fig1}
\end{figure}

Differential magnitudes on the instrumental system, which is close to Cousins I but was not transformed to it (since we had no color data) were computed for all stars by identifying a group of non-variable comparison stars. This was done by simply choosing the brightest few stars and averaging their magnitudes, taking care not to include any known association members or field variable stars. Variables reveal themselves by their larger than expected scatter relative to the comparison set during this process and may easily be weeded out. Only one such star was found in our four fields. Using this technique we found that the average error of a single photometric measurement was about 0.005 mag for bright stars and deteriorates to about 0.1 mag near the measuring limit. In Fig. \ref{fig1} we show the standard deviation ($\sigma$) computed for each star over the season in each field versus its average instrumental magnitude. Note that it is not possible to combine data from different fields because we have no absolute calibration of our instrumental magnitudes. However, all of the fields show the same expected form for the errors and it is rather easy to identify true variables on these figures.

Variable stars may be identified on these plots by their location significantly above the line defined by the bulk of the field stars. It is clear that seven of the nine program stars (plotted as diamonds on Fig. \ref{fig1}) are variable stars, while two are not. These two lie among the sequences defined by the field stars. We have also found one bright field star (plotted as a square in the figure) which is clearly variable (and periodic, as is shown below). One other bright field star, the brightest object in field C, appears to have a large value of $\sigma$ for its brightness. However, it is the brightest star in our sample by about a magnitude and turns out to be close to the saturation limit of the CCD chip near the peak of its point spread function on nights of good seeing. We attribute the star's apparent variability to this phenomenon and regard it here as likely to be a non-variable. There is also no indication of periodicity in its variations. In Table \ref{tab1} we provide data and results on the nine known members of MBM 12A which were within the boundaries of these fields and on the one variable ``field" star. Data in the first six columns are taken from \citet{l01}. The seventh column indicates whether the star can be identified as clearly variable based on it location on the plots shown in Fig. \ref{fig1}. The last four columns give the values of $\sigma$, full range of variation (in mag), period of the variation if it is identified as cyclic and false alarm probability (see next section). 

The variable field star that turned up by chance in our photometry has H$\alpha$ in absorption, so it has not been considered a candidate member of MBM 12A. However, it has been identified as a probable X-ray source by \citet{h00a}; it is star 38 in their study. By request, Hearty (private communication) re-examined his spectrum of the star and discovered an error in the published data. The correct spectral type (K3) and equivalent width of H$\alpha$ are given in our Table \ref{tab1}. The star does not have a detectable Li absorption line. It is undoubtedly a rather young star, but not young enough to be considered a potential member of MBM 12A, as we discuss further below.

The two association members (MBM 12A-7 and MBM 12A-11) that were not detected as variable are also the two faintest and latest-type (M6) stars in this sample. Both factors could be involved in our lack of detection of variability. Obviously, the increasing importance of measurement errors as the stars become fainter makes it increasingly difficult to identify true variables as stellar brightness declines. Also, since it is presumably cool spots which produce the variations in most cases, the increasing lack of contrast between the extremely cool stellar photospheres of these M6 stars and their spot regions, if they have any, might make it more difficult to detect rotation-induced variations. Further observations with better signal-to-noise or at more favorable wavelengths are obviously required to reveal any small photometric variations that may be present in these stars.

To summarize this section, seven of the nine previously identified members are found to be variable stars in this study. This reinforces the view that photometric variability can be a very useful criterion in identifying members of young stellar groups. Of course, the numbers in this association are small but, where they are larger (e.g. IC 348, ONC and NGC 2264), the result is the same. Virtually all pre-main sequence stars in clusters (at least earlier than M6) are variable stars at detectable levels (a few hundredths of a mag. or more) and it appears that the stars in this loose association are no different from their cluster counterparts in that regard. 

\section{Periodicity}

The periodogram technique of \citet{s82}, as formulated by \citet{hb} was used to search for periodicity in our data. We did this for every star measured, not just for known members of the association or obvious variable stars. The reason is that the periodogram technique is actually much more sensitive to true variability (if it is periodic) than the method of simply calculating a standard deviation since, for periodic stars, the signal is concentrated into a narrow band in frequency space while the noise is not. In other words, it is easier to detect low amplitude variability when it is periodic than when it is stochastic. As it turns out we did not, in fact, discover any star in these fields as being periodic which had not already revealed itself as a variable. However, by not limiting our photometry to known association members we did, at least,  discover that the variable field star is indeed periodic.

Periodogram functions for the nine association members and the one variable field star are shown in Fig. 2. Four of these ten stars show what we regard as significant peaks in their power spectra indicating the presence of true periodicity. These are marked on the figure and the periods (in days) given there and in Table \ref{tab1}. Based on their power levels one can estimate a false alarm probability (FAP) following the prescription given by \citet{hb}. As they discuss, however, their formulation only applies if each data point is independent of the others. This is not true in our original time series since we have more than one measurement at a closely spaced interval on some nights. To use the Horne \& Baliunas formula we must, therefore, average data obtained within the same night. The FAP values calculated by this means are given in the last column of Table \ref{tab1} for the stars identified as periodic. In all four cases, the chance of finding a peak in the periodogram of the given height in a random data set is less than 0.03, and in three cases it is less than 0.01. 

One star (MBM 12A-5) with a rather significant peak (FAP = 0.02) is not included in our list of positive detections because its period is  0.505 days. If this star is truly periodic with that period it would be impossible for us to discern it with confidence based on the current observations. The reason is that they were not frequent enough during any particular night to clearly distinguish such a putative cycle from an harmonic of the typical sampling interval of 1 day. Future observations with a higher cadence may reveal whether this star is, in fact, a cyclic variable with a period near 0.5 d. Another possibly periodic star (MBM12A-2), with FAP = 0.07, is noted in the table and discussed further in the next section.

It may be noted that all of the stars identified as periodic have more than one significant peak in their periodograms. The reason is quite simple: nightly observations from a single longitude introduce a sampling frequency of about one day into the data. When a truly periodic star is observed, the actual period ``beats" with the sampling interval to create an alias or ``beat" period. One can easily see that there are complementary peaks in the power spectra of all of these stars which are separated in frequency space by 1 $\pm$ 1/P where P is the true period. Identifying which is the true period and which is the beat period is not always easy, but there is little ambiguity for the four stars in this sample. In three cases, one of the peaks is significantly higher and the light curve phased with that period significantly less scattered than the other. In the case of MBM12A-4 there are two peaks of nearly equal height but the light curve phased with a period of 2.603 days looks better than the one phased at 0.722 days. Of course, it is possible that variations in light curve shape during an observing season could cause us to mistake the beat period for the true period but in samples where this can be tested (e.g. the ONC) it happens less than 15\% of the time, and usually with less well-defined light curves than are seen here \citep{h00}.  We conclude that the periods of these four stars are known with reasonable confidence, although v sin i measurements might be helpful in confirming them. 

Phased light curves for the four periodic stars are shown in Fig. \ref{fig3}. In all respects, the three known association members are representative in their properties of what is found for T Tauri stars in young clusters and associations such as Tau/Aur, IC 348, NGC 2264 and the ONC. In particular, the periods, amplitudes and shapes of the light curves are quite typical. In MBM 12A we find periods of 2.6 and 6.2 days for the two periodic K stars. The more populous clusters such as the ONC and NGC 2264 show peaks in the frequency distribution of rotation periods for these more massive stars (corresponding to spectral types earlier than M2) at around 1 and 4 days (NGC 2264) or around 2 and 8 days (ONC), as \citet{lm03} have shown. Lower mass stars (with spectral types later than M2) rotate faster in both clusters, and the period of the one mid-M star detected here (1.2 days) is in agreement with that trend. It is also numerically representative of values found in the extremely young clusters for such very low mass stars. If the possible 0.5 day period for MBM 12A-5 (a K star) were confirmed, that would be somewhat unusual, although perhaps the star could be understood as a slightly older, contracted and spun-up version of what is found in IC 348 and other extremely young clusters (see next section). 

Typical amplitudes for stars in extremely young clusters are several hundredths to a few tenths of a magnitude in Cousins I, similar to what is seen here. This is somewhat larger than is characteristic of older (30-100 My) clusters, where amplitudes above a few hundredths of a mag are unusual \citep{bsps}. It appears that the spots are simply larger or cooler on the extremely young stars, possibly reflecting a gradual decay in magnetic activity with age (in spite of their expected spin-up with contraction). It is also possible that the younger stars are simply more likely to maintain extremely non-axisymmetric spot distributions which give rise to their larger amplitudes. 

The situation is illustrated more quantitatively in Fig. \ref{fig4} where data from a $\sim$3 My old cluster, IC 348 \citep{chw} and from a $\sim$30 My old cluster, IC 2602 \citep{bsps} are compared. The V magnitude data for IC 2602 has been transformed to I assuming a typical ratio of amplitudes (in the sense of I to V) of 0.7 for spotted variables from \citet{h94} and full ranges are plotted, as opposed to half-ranges or amplitudes. It is clear that stars of different age occupy distinctly different, although adjacent, regions of this diagram. It is also clear that the MBM 12A stars with definite rotation periods (large squares on the figure) resemble the IC 348 stars much more so than they resemble the IC 2602 stars. Note that the fastest rotator in MBM 12A is an M5 star, which is of later spectral class than any star in IC 348 with a known rotation period. Since lower mass (M2 and later) stars in the ONC and in NGC 2264 are known to spin faster, as a group, than their more massive (K-M2 type) counterparts \citep{h01, lm03}, the location of this star to the left of the IC 348 cluster sequence is not surprising. Periods for such low mass stars in IC 348 are not yet available because the cluster, although comparable in distance to MBM 12, is more highly reddened and its low mass stars are below the magnitude limit for period detection at Van Vleck Observatory.

It is also interesting to find that the field X-ray source, RX J0255.9+2005, lies among the IC 2602 stars on the plot in Fig. \ref{fig4}. Its photometric properties are, therefore, consistent with the spectral data discussed above and given in Table \ref{tab1}. This is apparently a fairly young, but not extremely young star probably comparable in age to IC 2602 which is unlikely to be associated with MBM 12. To summarize, the MBM 12A stars do seem to be quite similar in period and amplitude to the other extremely young clusters we have monitored. In general, our periodic sample (small though it is) is much more representative of clusters of 1-5 My age such as IC 348, the ONC and NGC 2264 than it is of clusters of 30-100 My such as IC 2602 or the Pleiades. The periodic field star, on the other hand, has properties consistent with a significantly older population.

\section{The Large Amplitude Variables}

It is unusual to find WTTS with ranges exceeding $\sim$0.5 mag in Cousins I, although  V410 Tau is one exception \citep{h94}. It is also unusual to find periodicity in large amplitude CTTS although, again, there are exceptions. It is, therefore, somewhat surprising to find two large amplitude variables in this small sample, both of which {\it may} be periodic. The case of the WTTS MBM 12A-5 was mentioned briefly in the previous section, where it was noted as a possible cyclic variable with a period close to 0.5 days. This could be an analog of V410 Tau, which is a relatively rapidly rotating WTTS (period $\sim$ 1.8 days) that appears to be viewed close to edge on \citep{h89,s03}.  The other alternative is that the star is actually an accreting CTTS which happened to be observed in a low accretion state. It has recently been claimed \citep{lnhrt} that there are T Tauri stars which move from WTTS to CTTS. It is interesting that the star appears to have a K-L excess \citep{l01} but is also a close binary with the companions differing by only 0.5 mag in K \citep{c02}. It appears that this star will remain a bit of a puzzle until further observations are obtained. 

The other large amplitude variable is the CTTS MBM 12A-2 (LkH$\alpha$ 262), which was by far the most photometrically active star in our sample during the monitoring epoch. A light curve for the star is shown in Fig. \ref{fig5}. It is representative of the irregular light curves often displayed by CTTS. In particular, there is an epoch of relative stability (the early part of the observing season) followed by an epoch of more erratic behavior. There are both high frequency oscillations and slower drifts of mean brightness. The amplitude, 1.14 mag in Cousins I, is large but not highly unusual for a CTTS \citep{h94}. Normally, this type of variation (Type II) is attributed to the rotational modulation of hot accretion spots or zones (rings?) which vary due to unsteady accretion on typical time scales of days or less. Since the spots are usually not stable for as long as one rotation period, the variations are normally irregular. Sometimes they may be stable for up to a few rotation periods so that a period may be found in the photometry. This is referred to as Type IIp (for periodic) variability.

It is possible that MBM12A-2 is a Type IIp variable. Its periodigram (Fig. 2) shows a fairly high peak at a period of 3.13 days. The false alarm probability is 7\% so it does not rise to the usual level that we would accept for definite periodicity. Since the light curve changed its basic appearance between the beginning and end of the monitoring time we searched for periods independently within these two time intervals. Nothing more significant was found by this process. As is typical of light curves with normalized powers of only a little more than 6, there is considerable scatter and the phased light curve does not look convincing. In a sample of 10 stars we would expect one to show a false alarm period of about this light curve quality, so the variations may actually be truly stochastic. On the other hand, since Type II variables can change rapidly from cycle to cycle the period could be real but hard to see because of the ``noise" injected by the the variations in the spots. Unfortunately, there is no clear conclusion that can be reached on the basis of the present data. We can only say that the star is a large amplitude variable typical of CTTS in other young stellar groups and that it may have a period of 3.13 days. This is the type of star that would benefit from more intensive monitoring than is possible at Wesleyan due to the inevitable interruptions caused by clouds.

The rather low amplitude (non-periodic) variability of the other CTTS in MBM 12A (star 3 = LkH$\alpha$ 263) is not surprising. Many CTTS show only small amounts of variability at certain times (or ever). It is well known that there is no tight correlation between H$\alpha$ equivalent width and amplitude of variability for a set of young stars \citep{h94} and none is seen here. For particular stars with Type II characteristics there is a good correlation between brightness in the V band and H$\alpha$ flux \citep{bb, h94} but there is no correlation that extends across a group. It is unclear whether the non-simultaneity of the photometric and H$\alpha$ measurements contributes to the lack of a correlation in the present data. As Table \ref{tab1} shows, there can be quite a range of H$\alpha$ equivalent widths measured for the same star at different times. Unless values appropriate to the time of photometric monitoring are used, it is probable that any existing correlation would be hidden by the effects of non-simultaneity. 

To summarize this section, we find two large amplitude variables in MBM 12A both of which are somewhat unusual. MBM 12A-5 is a WTTS which possibly has a very short period and could be a V410 Tau analog. It is unusual to find such a large amplitude for a WTTS. MBM 12A-2 is a CTTS and has the variability characteristics of a Type II star, but has a possible period, which would make it a Type IIp star. These are fairly rare. In IC 348, for example, \citet{chw} found rotation periods for more than two dozen WTTS, none of which had large amplitudes, but did not find periods for any of the  20 CTTS in their monitored field. A more complete comparison of the variability properties of IC 348 and MBM 12A is interesting and we turn to that in the final section of the paper.

\section{Discussion and Summary}

We may compare the variability characteristics of the MBM 12A stars, as a group, to those of the well known and well studied young cluster IC 348, which has been monitored photometrically at Wesleyan for five years \citep{hmw,chw}. Since the distance (260-320 pc) and age ($\sim$1-3 My) of IC 348  \citep{h98,lrll,l03} are quite similar to MBM 12A, this is an interesting comparison to make. Limiting ourselves to the brightest stars (I $<$ 14.3) where, in both clusters, we have the most complete and sensitive survey of variability characteristics, and to K or M spectral classes, we find that 19/21 such stars are detected as variable in IC 348, compared to 7/7 in MBM 12A. Periodicity is found in 14/21 of the cluster stars and in 3/7 of the association members. Large amplitude ($>$0.5 mag) irregular variability is found in 2/21 of the cluster stars and 1/7 of the association members (possibly 2/7 if you count star 5, which may not actually be periodic). Even though the numbers are small, there is a clear sense that the overall variability properties of stars in the MBM 12A association compare well with those in IC 348, supporting the view that we are viewing stellar groups of about the same evolutionary status.  

Periods range from 2.2 to 16.4 d for the 14 bright K and M stars in IC 348, whereas they are 1.2,2.6 and 6.2 days for the three periodic stars in MBM 12A. This is marginally surprising and would be a more significant difference if MBM 12A-5 is confirmed to have a period as short as 0.5 d. In general, it would appear possible that the MBM 12A stars, as a group, do tend to spin a bit faster than the IC 348 stars. This would not be unexpected if there is an age difference. In fact, \citet{lm03} have recently shown that the stars in NGC 2264 spin faster (by about a factor two) than their ONC counterparts because they are a bit older. During the first few million years, radius (R) depends on age (t) for K and M-type PMS stars as R $\propto$ t$^{-{1 \over 3}}$, so if angular momentum is conserved one finds P $\propto$ R$^2$ $\propto$ t$^{-{2 \over 3}}$ \citep{ch,hbj}. An age difference of a factor of three between IC 348 and MBM 12A in the sense that MBM 12A is older would lead, therefore, to shorter periods for the association stars by about a factor of 2 compared to the cluster, as may be suggested by our small data sample. 

There is one other small difference which may or may not be of importance between these groups and their variability characteristics. In IC 348 it is quite striking that every periodic star is a WTTS, not a CTTS. In MBM 12A, on the other hand, two of the three periodic stars have H$\alpha$ equivalent widths which qualify them as CTTS, even at the lowest points in their ranges. We have no idea whether this is significant or, if so, how it could arise and simply note it here as a curious outcome of our work. Perhaps related to it is the also curious fact that in this MBM 12A sample 4/7 stars are CTTS, whereas in the corresponding bright IC 348 sample the fraction of CTTS is 5/21. One purely speculative cause of this could be that disks generally survive longer in loose associations than in denser clusters.

To summarize, we find that every bright member of MBM 12A is a photometric variable and that three of them are definitely periodic. The photometric properties of this small sample of association stars are, in most respects, indistinguishable from stars in the extremely young cluster IC 348, suggesting a similar age of $\sim$few My. The periodic stars occupy a clearly different region of  period-amplitude space than an older comparison group (IC 2602). We also find one CTTS (LkH$\alpha$ 262) which has a large amplitude, and one possibly cyclic variable of large amplitude and very short period (MBM12A-5). With such a small sample of stars it is impossible to make finer distinctions with certainty but the indications are that, if anything, the MBM 12A stars spin slightly faster than the IC 348 stars suggesting they might be a bit older, perhaps by as much as a factor of three. Clearly, however, based on their photometric properties alone they are in the 1-10 My range and not as old as 30 My. A field periodic variable discovered serendipitously, which is associated with an X-ray object (RX J0255.9+2005), appears to be a representative of an older (perhaps 30-100 My) population in the same direction and is unlikely to be associated with the MBM 12 dark cloud complex. 

\acknowledgments

We thank the referee, J. Stauffer, for his helpful report and K. Luhman for additional useful comments on the first draft of this manuscript. We also thank T. Hearty for the revised data on RX J0255.9+2005. We thank the W. M. Keck Foundation for their support of the Keck Northeast Astronomy Consortium. It was under its auspices that W. Hawley spent the summer of 2003 at Wesleyan working on this project. We thank the large number of Wesleyan students who contributed to this project by obtaining the data on our campus telescope. This investigation was supported by grants from the NASA-Origins program.

\clearpage

\begin{figure}
\plotone{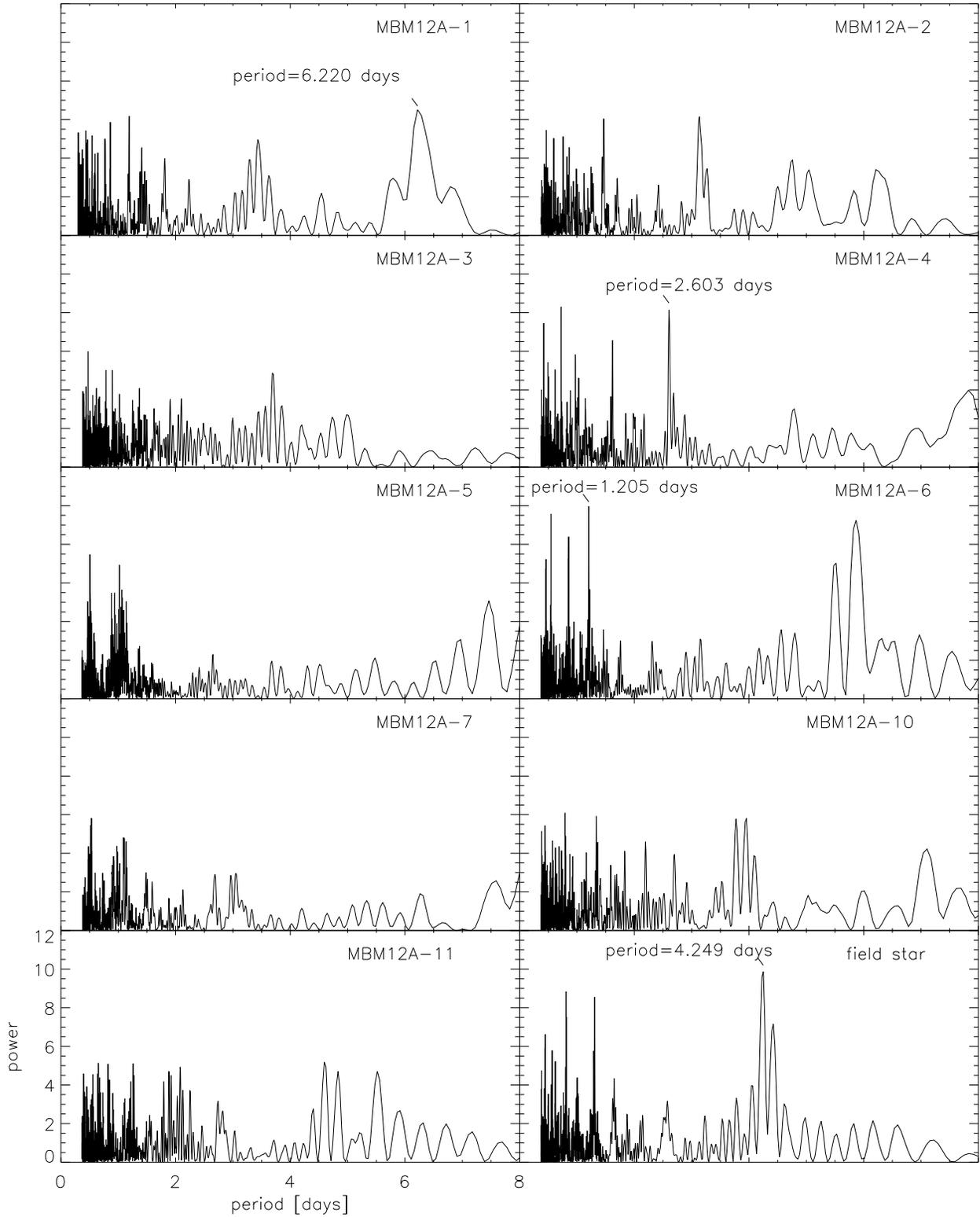}
\label{figure2}
\caption{Periodograms, based on the Scargle (1982) technique are shown for the nine members of MBM 12A and for the one variable field star. (cont. next page)}
\end{figure}
\clearpage

Fig. 2. (cont.) In four cases, the power is 
sufficiently high that the existence of a true period is quite certain, and these are indicated on the figures. One star (MBM 12A-5) with rather high power at a period of 0.5 days was deemed not believable on the basis of the similarity of the period to an harmonic of a sampling interval. Another star (MBM 12A-2) has a marginally significant peak at a period of 3.13 days. 

\clearpage
\begin{figure}
\plotone{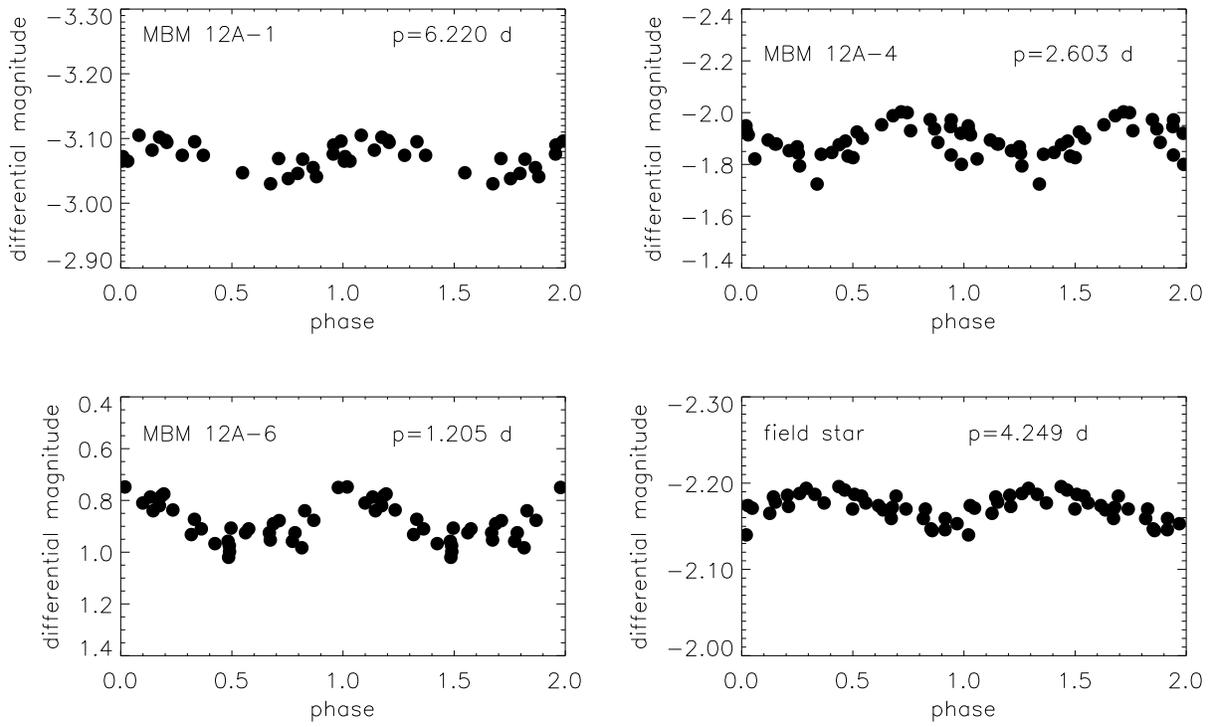}
\caption{Light curves for the four periodic variables phased with their adopted periods.}
\label{fig3}
\end{figure}
\clearpage

\begin{figure}
\plotone{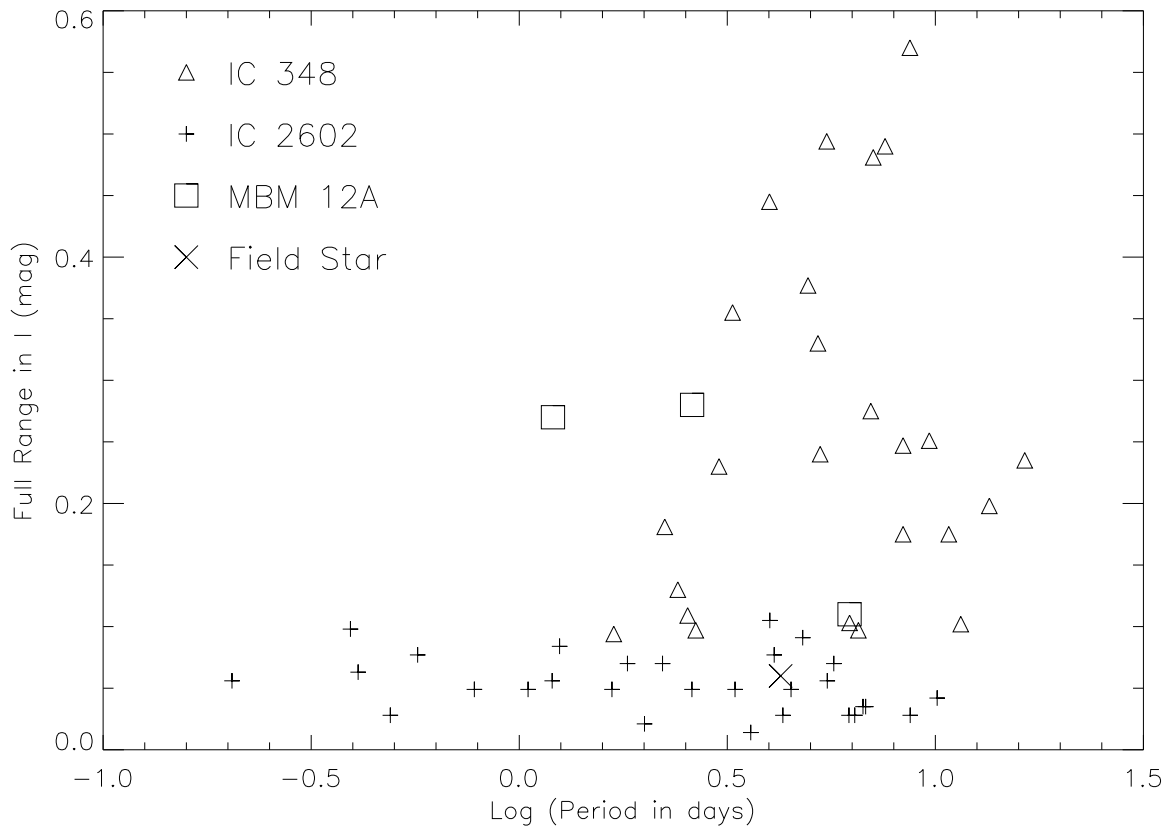}
\caption{The full amplitude in Cousins I (mag) is plotted versus the logarithm of the rotation period (in days) for periodic variables in young clusters and in MBM 12A (large squares). Note that the shortest period star in MBM 12A has a later spectral type (M5) than any star in IC 348 and this fact may account for its somewhat extreme position. The periodic field star associated with the X-ray source RX J0255.9+2005 is also shown as a large "X". The data for IC 348, which has an estimated age of $\sim$1-3 My, is from \citet{chw}. The data for IC 2602, which has an estimated age of $\sim$30 My, is from \citet{bsps}. Clearly, the MBM 12A periodic variables, as a group, resemble stars in IC 348 while the field X-ray object more resembles stars in IC 2602.}
\label{fig4}
\end{figure}
\clearpage

\begin{figure}
\plotone{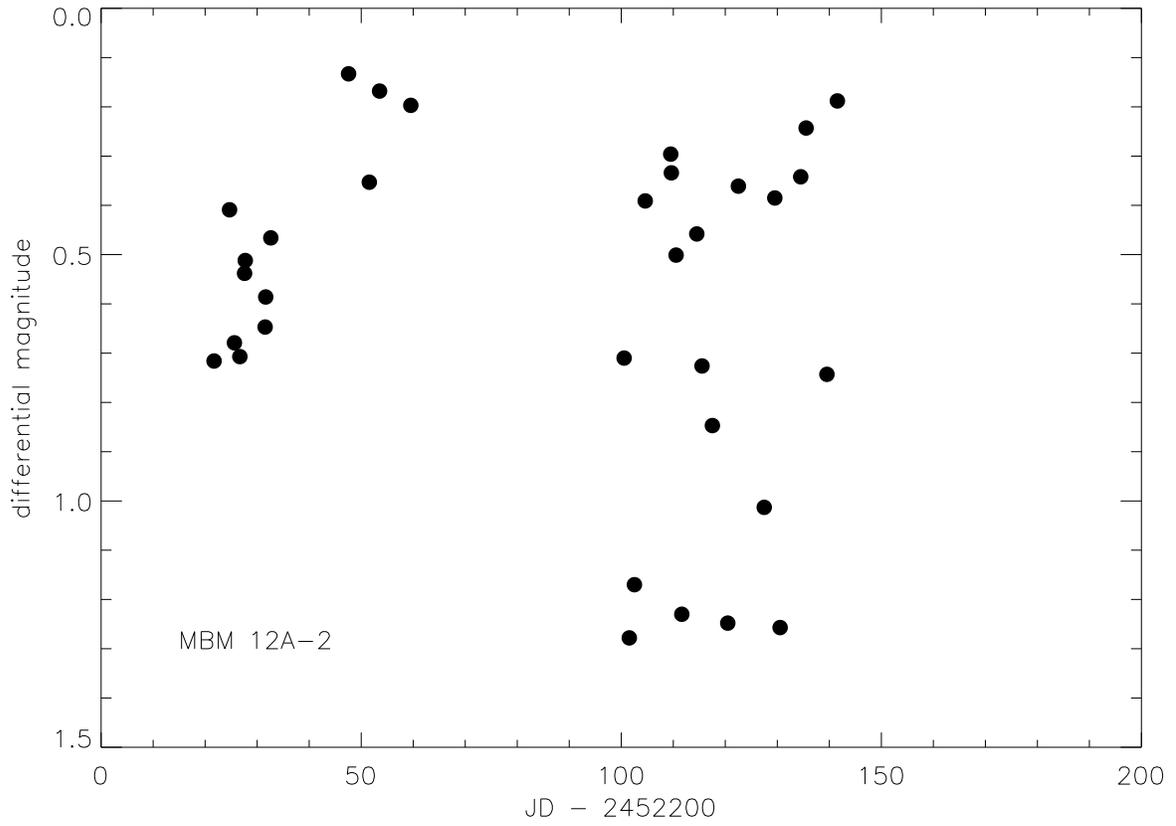}
\caption{The light curve of MBM 12A-2 (LkH$\alpha$ 262), the largest amplitude variable in our sample. It is possibly a cyclic variable with a period of 3.13 days (see Fig. 2).}
\label{fig5}
\end{figure}
\clearpage

\begin{deluxetable}{ccccccccccc}
\tablenum{1}
\tabletypesize{\scriptsize}
\tablecaption{Results of Photometric Monitoring of MBM 12A Stars \label{tab1}}
\tablewidth{0pt}
\tablehead{\colhead{Star$^*$} & \colhead{Name} & \colhead{Field} & \colhead{I (mag)}   & 
\colhead{SpT} & \colhead{EW(H-alpha)} & \colhead{var?}   & 
\colhead{Sigma} & \colhead{Range}   & 
\colhead{Period} &  \colhead{FAP$^*$}}
\startdata
1 & RX J0255.4 +2005 & A & 10.69 & K6 & -1.1 & yes & 0.035 & 0.11 & 6.22 & 0.03\\
2 & LkH$\alpha$ 262 & B & 12.51 & M0 & -32 to -40 & yes & 0.342  & 1.14  & 3.13? & 0.07 \\
3 & LkH$\alpha$ 263 & B & 12.17 & M3 & -13 to -33 & yes & 0.031 & 0.12 &  & \\
4 & LkH$\alpha$ 264 & B & 11.60 & K5 & -17 to -59 & yes & 0.065 & 0.28 & 2.603 & 0.007\\ 
5 & E 02553 +2018 & C & 10.54 & K4 & -2.5 & yes & 0.126 & 0.56 & 0.505? & 0.02 \\
6 & RX J0258.3 +1947 & D & 13.40 & M5 & -25 to -34 & yes & 0.076 & 0.27 & 1.205 & 0.002 \\
7 & RX J0256.3 +2005 & B & 14.70 & M6 & -13.5 & no & 0.013 & 0.06 &  & \\
10 & & C & 13.04 & M3 & -12 & yes & 0.098 & 0.35 &  & \\
11 & & D & 15.30 & M6 & -13.5 & no & 0.021 & 0.11 & &  \\
field$^*$ & RX J0255.9+2005 & A & (11.6) & K3  & +1.4 & yes & 0.015 & 0.06 & 4.249 & 0.006 \\
\enddata
\tablenotetext{*}{Star numbers are from Luhman (2001). FAP is the false alarm probability as described in the text. The field star is star 38 in Hearty et al. (2000a). It is probably associated with their detected X-ray source RX J0255.9+2005. Hearty (private communication) indicates that the published spectral data for this star in table 4 of Hearty et al. (2000a) is incorrect. The correct data is given here. Also, the star does not have detectable Li absorption.  The I magnitude given is an estimate based on the difference between it and star 1.}
\end{deluxetable}

\end{document}